\documentclass[%
 reprint,
 amsmath,amssymb,
 aps,
 pre,
 longbibliography
]{revtex4-2}

\usepackage{graphicx}
\usepackage{dcolumn}
\usepackage{bm}
\usepackage[T1]{fontenc}
\usepackage{hyperref}
\hypersetup{
colorlinks = true,
linkcolor = blue,
citecolor = blue
}
\usepackage{here} 

\usepackage[version=4]{mhchem}
\allowdisplaybreaks 
\usepackage{color}

\begin{document}


\title{Information Propagation in Predator-Prey Dynamics of Turbulent Plasma}

\author{Tomohiro Tanogami$^1$}
\author{Makoto Sasaki$^{2}$}
\author{Tatsuya Kobayashi$^{3,4}$}
\affiliation{$^1$Department of Earth and Space Science, The University of Osaka, Osaka 560-0043, Japan\\
$^2$College of Industrial Technology, Nihon University, Narashino 275-8575, Japan\\
$^3$National Institute for Fusion Science, National Institutes of National Sciences, Toki 509-5292, Japan\\
$^4$The Graduate University for Advanced Studies, SOKENDAI, Toki 509-5292, Japan}




\date{\today}
\begin{abstract}
Magnetically confined fusion plasmas exhibit predator-prey-like cyclic oscillations through the self-regulating interaction between drift-wave turbulence and zonal flow.
To elucidate the detailed mechanism and causality underlying this phenomenon, we construct a simple stochastic predator-prey model that incorporates intrinsic fluctuations and analyze its statistical properties from an information-theoretic perspective.
We first show that the model exhibits persistent fluctuating cyclic oscillations called quasi-cycles due to amplification of intrinsic noise.
This result suggests the possibility that the previously observed periodic oscillations in a toroidal plasma are not limit cycles but quasi-cycles, and that such quasi-cycles may be widely observed under various conditions.
For this model, we further prove that information of the zonal flow is propagated to turbulence.
This result suggests that turbulence behavior may be predictable to a certain extent based on zonal flow characteristics.
\end{abstract}

\pacs{Valid PACS appear here}

\maketitle
\section{Introduction}
Predator-prey dynamics can be observed ubiquitously in various systems such as ecosystems~\cite{may1973stability,maynard1974models,murray2007mathematical}, genetic systems~\cite{xue2016stochastic}, fluids~\cite{shih2016ecological,shih2017spatial,goldenfeld2017turbulence,wang2022stochastic}, or even quantum systems~\cite{PhysRevX.15.011010}.
Magnetically confined fusion plasmas also exhibit predator-prey-like cyclic oscillations in various situations, including the self-regulating process between zonal flow and drift-wave turbulence~\cite{diamond2005zonal,itoh2006physics,itoh2013assessment,gurcan2015zonal}, the intermediate phase (I-phase) during the low-to-high confinement (L-H) transition~\cite{estrada2010experimental,conway2011mean,schmitz2012role,cheng2013dynamics,kobayashi2013spatiotemporal}, the formation and collapse of internal transport barriers~\cite{bizarro2007controlling}, and self-regulated oscillations of magnetic islands~\cite{kobayashi2022self}.
Among these, the interplay between zonal flow (predator) and drift-wave turbulence (prey) is believed to play a crucial role in suppressing anomalous heat and particle transport, and therefore, intensive attempts have been made to regulate turbulence by controlling zonal flow.
Despite this importance, the detailed mechanism and causality underlying the cyclic oscillations have not been elucidated~\cite{itoh2013assessment}.
To deepen our understanding of the self-regulating process between zonal flow and turbulence, it is desirable to construct a minimal model that exhibits the cyclic oscillations and investigate the causality within the oscillations.

The main aim of this Letter is twofold.
The first is to construct a simple qualitative model for the cyclic oscillations observed in plasma turbulence.
Note that the standard two-variable predator-prey model proposed by Diamond \textit{et al.}~\cite{diamond1994self,diamond2005zonal}, which incorporates the self-regulation mechanism between turbulence and zonal flow, does not exhibit a stable cyclic oscillation.
To address this issue, we aim to construct a stochastic predator-prey model that incorporates the effects of intrinsic noise from the perspective of statistical physics.
Since it has been pointed out in the field of ecology that it is essential to consider intrinsic fluctuations in predator-prey dynamics~\cite{mckane2005predator,black2012stochastic}, we expect that such a statistical physics approach provides valuable insights into plasma turbulence.
The second is to quantify the predictive causality between zonal flow and turbulence in this model through the lens of information theory.
Several previous studies have attempted to quantify the predictive causality in plasma turbulence using information-theoretic quantities, such as (net) transfer entropy~\cite{schreiber2000measuring,van2014causality} or information length~\cite{kim2020time,hollerbach2020time,kim2022non,kim2023stochastic,fuller2023time,fuller2024time,kim2024nonperturbative,kim2025probabilistic}.
However, because these quantities are not antisymmetric under time reversal, it is unclear whether they appropriately quantify predictive causality.
Instead, here we employ \textit{information flow}, which is a pivotal concept within the framework of information thermodynamics~\cite{allahverdyan2009thermodynamic,horowitz2014thermodynamics,parrondo2015thermodynamics,chetrite2019information,parrondo2015thermodynamics,peliti2021stochastic,shiraishi2023introduction} and has recently been applied to various systems, including standard fluid turbulence~\cite{tanogami2024information,tanogami2024scale,tanogami2025amplify}.
Because the information flow is antisymmetric under time reversal and becomes zero in an equilibrium state where all probability currents are absent, it may reflect some aspects of the predictive causality consistent with the behavior of the probability currents.

The constructed model can be interpreted as a standard two-variable predator-prey model that takes into account intrinsic fluctuations.
We show that the model exhibits persistent fluctuating cyclic oscillations, called \textit{quasi-cycles}~\cite{mckane2005predator,black2012stochastic}, due to stochastic amplification.
Quasi-cycles are distinguished from both deterministic and stochastic limit cycles by their shorter correlation time scales~\cite{pineda2007tale,butler2011fluctuation}. 
Furthermore, they can readily emerge in any system possessing a stable spiral fixed point subject to noise.
This result suggests the possibility that the previously observed periodic oscillations in a toroidal plasma are not limit cycles, but rather quasi-cycles.
Our analysis of the resonance condition also implies that such quasi-cycles may be widely observed under various conditions.

For this model, we quantify the causality between zonal flow and turbulence using information flow.
Specifically, we analytically show that the information on zonal flow is propagated to turbulence.
This information propagation occurs as long as zonal flow and turbulence coexist, regardless of the presence or absence of quasi-cycles.
Our finding suggests that turbulence behavior may be predictable to a certain extent based on zonal flow characteristics.

\begin{figure}[t]
\center
\includegraphics[width=8.6cm]{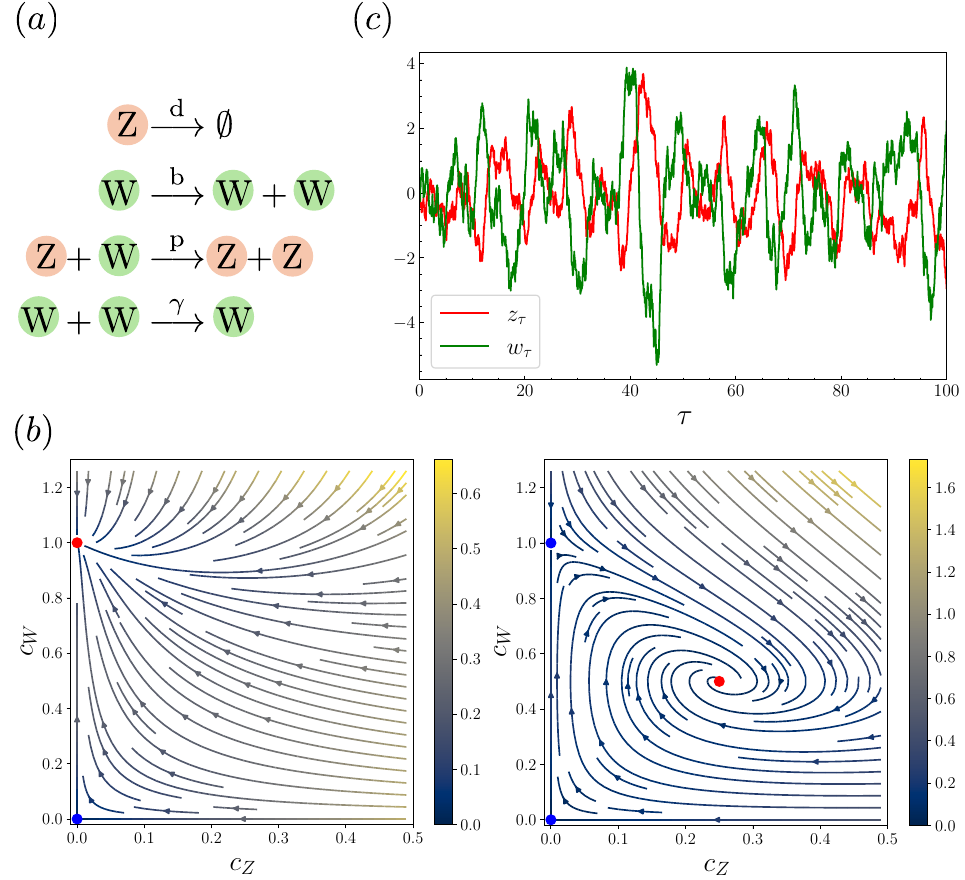}
\caption{(a) Schematic of the model. 
(b) Vector fields for parameter values $p=0.5, d=b=\gamma=1$ (left) and $p=2, d=b=\gamma=1$ (right). The color bar denotes the magnitude of the vector fields. The red markers indicate the stable fixed points, and the blue markers indicate the unstable fixed points.
(c) The trajectory of the Langevin equation. The parameter values are $\epsilon=1/2, \beta=1, D^W=1/2$. The characteristic period is given by $2\pi/\omega_c\simeq10$.
}
\label{fig:four_pictures}
\end{figure}

\section{Setup}
Our model can be expressed using chemical reaction equations for two types of ``reactants'': zonal flow (predator) $\ce{Z}$ and drift-wave turbulence (prey) $\ce{W}$. 
Each of these reactants $\ce{Z}$ and $\ce{W}$ may be interpreted as representing arbitrary energy levels that are much smaller than their characteristic large-scale energies but are large enough for the following reaction equations to hold.
There are four ``reactions'' in our model (Fig.~\ref{fig:four_pictures}(a)):
the linear damping of the zonal flow, $\ce{Z ->[d] \emptyset}$, where $\ce{\emptyset}$ denotes the null state;
the linear growth of the turbulence, $\ce{W ->[b] 2W}$;
the suppression of the turbulence by the zonal shear, $\ce{Z + W ->[p] 2Z}$;
the nonlinear saturation of the turbulence, $\ce{2W ->[\gamma] W}$.
Here, $d$ denotes the damping rate, $b$ denotes the linear growth (birth) rate, $p$ denotes the suppression (predation) rate, and $\gamma$ denotes the nonlinear saturation rate.
Note that these reactions incorporate a minimal self-regulation mechanism in which turbulence $\ce{W}$ drives zonal flow $\ce{Z}$, which in turn suppresses turbulence.
Below, we use the notation $\rho\in\{d,b,p,\gamma\}$ to represent both the rate constant and the label for each reaction.
Let ${\bm n}_t=(n^Z_t,n^W_t)\in\mathbb{Z}^2_+$ be the total energies of \ce{Z} and \ce{W} at time $t$ and $p_t({\bm n})$ be probability of finding ${\bm n}_t={\bm n}$.
Then, the master equation for this model reads
\begin{align}
\partial_tp_t({\bm n})&=\sum_\rho\left[W_\rho({\bm n}|{\bm n}-{\bm S}_\rho)p_t({\bm n}-{\bm S}_\rho)\right.\notag\\
&\qquad\qquad\qquad\left.-W_\rho({\bm n}+{\bm S}_\rho|{\bm n})p_t({\bm n})\right].
\label{master eq}
\end{align}
Here, $W_\rho({\bm n}+{\bm S}_\rho|{\bm n})$ denotes the transition rate from state ${\bm n}$ to state ${\bm n}+{\bm S}_\rho$ due to the reaction $\rho$, where ${\bm S}_\rho=(S^Z_\rho,S^W_\rho)$ denotes the stoichiometric vectors, which quantifies the change of ${\bm n}$ during a reaction of type $\rho$.
For simplicity, we assume the mass action law~\cite{rao2018conservation}:
\begin{align}
W_\rho({\bm n}+{\bm S}_\rho|{\bm n})&=\Omega \rho\prod_{\alpha\in\{Z,W\}}\dfrac{1}{\Omega^{\nu^\alpha_\rho}}\dfrac{n^\alpha!}{(n^\alpha-\nu^\alpha_\rho)!},
\end{align}
where the stoichiometric coefficients ${\bm \nu}_\rho=(\nu^Z_\rho,\nu^W_\rho)$ denote the number of reactants \ce{Z} and \ce{W} involved in each reaction $\rho$, and $\Omega$ denotes the system size, which may be interpreted as a total volume of the plasma.
Although we can estimate parameter values that are consistent with experiments and simulations, we consider this model as a qualitative description and will not explore its quantitative aspects in depth.
We note that similar models have been used for the laminar-turbulence transition in standard fluids to investigate the connection with the directed percolation universality class~\cite{shih2016ecological,shih2017spatial,goldenfeld2017turbulence,wang2022stochastic}.

\section{System size expansion}
Since we are interested in the collective behavior of the system, we consider coarse-grained descriptions by taking the infinite volume limit $\Omega\rightarrow\infty$.
Here, we note that the two limits $t\rightarrow\infty$ and $\Omega\rightarrow\infty$ do not commute in this model, which is known as Keizer's paradox~\cite{vellela2007quasistationary,falasco2023macroscopic}.
Indeed, if we take the limit $t\rightarrow\infty$ first, then the probability distribution converges to a unique stationary distribution $p_{\mathrm{ss}}({\bm n})=\delta_{{\bm n},{\bm 0}}$, which implies that eventually there will be no zonal flows and turbulences.
Below, we focus on the opposite case, where we first take $\Omega\rightarrow\infty$ and then $t\rightarrow\infty$.
The coarse-grained dynamics in the limit $\Omega\rightarrow\infty$ can be systematically obtained by applying the van Kampen system size expansion (also known as the linear noise approximation)~\cite{van1992stochastic}.
From the central limit theorem, we first note that $p_t({\bm n}/\Omega)$ can be reasonably approximated as a Gaussian with a finite width of order $\Omega^{-1/2}$.
To separate the fluctuating part from the average part, we introduce a new stochastic variable ${\bm r}_t=(z_t,w_t)\in\mathbb{R}^2$ as ${\bm n}_t/\Omega={\bm c}_t+{\bm r}_t/\Omega^{1/2}$, where ${\bm c}_t=(c^Z_t,c^W_t)\in\mathbb{R}^2_{\ge0}$ denotes the mean energy density to be determined.
By substituting this expression into Eq.~(\ref{master eq}) and expanding in terms of $\Omega^{-1/2}$, we can obtain the time evolution equations for ${\bm c}_t$ and ${\bm r}_t$, respectively.

\subsection{Deterministic rate equation}
The leading order $O(\Omega^{1/2})$ yields the time evolution equation for the average part ${\bm c}_t$, which can be regarded as a coarse-grained dynamics of Eq.~(\ref{master eq}):
\begin{align}
\dfrac{d}{dt}c^Z_t&=p c^Z_tc^W_t-d c^Z_t,\label{rate eq_z}\\
\dfrac{d}{dt}c^W_t&=b c^W_t-\gamma (c^W_t)^2-p c^Z_tc^W_t.\label{rate eq_w}
\end{align}
See Appendix~\ref{Detailed calculation of system size expansion} for a detailed derivation.
This equation corresponds to the standard two-variable predator-prey model proposed by Diamond \textit{et al.}~\cite{diamond1994self,diamond2005zonal}, which has the same form as the Lotka--Volterra equation.
There are two regimes in this model (Fig.~\ref{fig:four_pictures}(b)).
When $bp<d\gamma$, the stable fixed point is $(c^Z_{\mathrm{ss}},c^W_{\mathrm{ss}})=(0,b/\gamma)$, which represents a state without zonal flows. 
In contrast, when $bp>d\gamma$, the stable fixed point is $((bp-d\gamma)/p^2,d/p)$, which represents a coexistence state.
Importantly, this equation does not exhibit limit cycles, contrary to the naive expectation of the existence of predator-prey oscillations.
While cyclic oscillations emerge when $\gamma=0$, these cycles are not limit cycles but structurally unstable.

\subsection{Linear Langevin equation}
The coarse-grained description [Eqs.~(\ref{rate eq_z}) and (\ref{rate eq_w})] ignores intrinsic noise that is inevitably present in the original individual-level description (\ref{master eq}).
For a realistic situation where $\Omega$ is large but finite, it becomes essential to consider the intrinsic noise. 
To investigate the effect of the intrinsic noise on the coexistence state, we focus on the regime $bp>d\gamma$ and proceed to the subleading order $O(\Omega^0)$ of the system size expansion by substituting $(c^Z_{\mathrm{ss}},c^W_{\mathrm{ss}})=((bp-d\gamma)/p^2,d/p)$.
The resulting equation describes the time evolution of the fluctuating part ${\bm r}_t$ around ${\bm c}_t={\bm c}_{\mathrm{ss}}$ (see Appendix~\ref{Detailed calculation of system size expansion} for a detailed derivation):
\begin{align}
\dfrac{d}{d\tau}z_\tau&=\epsilon w_\tau+\zeta^Z_\tau,\label{linear Langevin_z}\\
\dfrac{d}{d\tau}w_\tau&=-\beta z_\tau-(1-\epsilon)w_\tau+\zeta^W_\tau,\label{linear Langevin_w}
\end{align}
where $\tau:=b t$ denotes the dimensionless time, $\beta:=d/b$ denotes the dimensionless damping rate, and $\epsilon:=1-d\gamma/bp\in(0,1)$ denotes the ratio of time scales for $z$ and $w$.
The terms $\zeta^Z_\tau$ and $\zeta^W_\tau$ denote the zero-mean white Gaussian noise that satisfies $\langle\zeta^Z_\tau\zeta^Z_{\tau'}\rangle=2D^Z\delta(\tau-\tau')$, $\langle\zeta^Z_\tau\zeta^W_{\tau'}\rangle=-D^Z\delta(\tau-\tau')$, and $\langle\zeta^W_\tau\zeta^W_{\tau'}\rangle=2D^W\delta(\tau-\tau')$, where $D^Z:=\epsilon c^W_{\mathrm{ss}}=\epsilon d/p$ and $D^W:=c^W_{\mathrm{ss}}=d/p$.

We emphasize that the statistical properties of the noise, such as the noise intensities $D^Z$ and $D^W$, are uniquely determined from the master equation (\ref{master eq}).
In other words, the noise driving the system is not artificially added but rather arises from intrinsic stochasticity.
This property contrasts with the stochastic models proposed in Refs.~\cite{kim2020time,hollerbach2020time,kim2023stochastic,fuller2023time,fuller2024time,kim2024nonperturbative,kim2025probabilistic}, where noise is artificially added to the corresponding deterministic predator-prey models, and its properties can be arbitrarily determined.
We also remark that the noises $\zeta^Z_\tau$ and $\zeta^W_\tau$ in our model are not independent but are correlated as $\langle\zeta^Z_\tau\zeta^W_{\tau'}\rangle=-D^Z\delta(\tau-\tau')$, which is a property not considered in previously proposed models.

Note that the linear Langevin equations (\ref{linear Langevin_z}) and (\ref{linear Langevin_w}) have a form in which noise is added to the linearized Lotka--Volterra equation around the coexistence fixed point. 
To see this point explicitly, we rewrite Eqs.~(\ref{linear Langevin_z}) and (\ref{linear Langevin_w}) in vector form with ${\bm r}_\tau=(z_\tau,w_\tau)$:
\begin{align}
\dfrac{d}{d\tau}{\bm r}_\tau=-\mathsf{A}{\bm r}_\tau+\mathsf{B}{\bm \xi}_\tau,
\end{align}
where
\begin{align}
\mathsf{A}=
\begin{pmatrix}
0 & -\epsilon \\
\beta & 1-\epsilon
\end{pmatrix},
\label{def A}
\end{align}
and ${\bm \xi}$ denotes the zero-mean white Gaussian noise that satisfies $\langle\xi^\alpha_\tau\xi^{\alpha'}_{\tau'}\rangle=\delta^{\alpha\alpha'}\delta(\tau-\tau')$ ($\alpha=Z,W$), which is multiplied by the noise matrix ${\mathsf{B}}$ that satisfies
\begin{align}
\mathsf{B}\mathsf{B}^\top=
\begin{pmatrix}
2D^Z & -D^Z \\
-D^Z & 2D^W
\end{pmatrix}.
\label{def B}
\end{align}
Then, it is easy to see that $\mathsf{A}$ corresponds to the coefficient matrix in the linearized Lotka--Volterra equation (\ref{rate eq_z}) and (\ref{rate eq_w}) at the coexistence fixed point.

The linear Langevin equations (\ref{linear Langevin_z}) and (\ref{linear Langevin_w}) are equivalent to the Fokker--Planck equation,
\begin{align}
\partial_\tau p_\tau({\bm r})=-\dfrac{\partial}{\partial z}J^Z_\tau({\bm r})-\dfrac{\partial}{\partial w}J^W_\tau({\bm r}),
\end{align}
where $p_\tau({\bm r})$ denotes the probability density of finding ${\bm r}_\tau={\bm r}$ and $J^Z_\tau({\bm r})$ and $J^W_\tau({\bm r})$ denote the probability currents defined by
\begin{align}
J^Z_\tau({\bm r})&:=\epsilon wp_\tau({\bm r})-D^Z\dfrac{\partial}{\partial z}p_\tau({\bm r})+\dfrac{D^Z}{2}\dfrac{\partial}{\partial w}p_\tau({\bm r}),\\
J^W_\tau({\bm r})&:=\bigl[-\beta z-(1-\epsilon)w\bigr]p_\tau({\bm r})\notag\\
&\qquad-D^W\dfrac{\partial}{\partial w}p_\tau({\bm r})+\dfrac{D^Z}{2}\dfrac{\partial}{\partial z}p_\tau({\bm r}).
\end{align}
Because the coefficient matrix $\mathsf{A}$ has eigenvalues $\lambda_\pm$ with strictly positive real parts, there exists a unique steady-state probability density $p_{\mathrm{ss}}({\bm r})$ for this Fokker--Planck equation~\cite{gardiner1985handbook,risken1996fokker}.
It can be explicitly calculated as
\begin{align}
p_{\mathrm{ss}}({\bm r})=\dfrac{1}{2\pi\sqrt{\det\Sigma}}\exp\left(-\dfrac{1}{2}({\bm r}-\langle{\bm r}\rangle)^\top\Sigma^{-1}({\bm r}-\langle{\bm r}\rangle)\right),
\label{Gaussian}
\end{align}
where $\Sigma:=\langle({\bm r}-\langle{\bm r}\rangle)({\bm r}-\langle{\bm r}\rangle)^\top\rangle$ denotes the covariance matrix and $\langle\cdot\rangle$ denotes the average with respect to this distribution.
Importantly, the steady state is generally out of equilibrium.
That is, the steady-state probability current does not vanish, and there exists ${\bm r}$ such that ${\bm J}_{\mathrm{ss}}({\bm r}):=(J^Z_{\mathrm{ss}}({\bm r}),J^W_{\mathrm{ss}}({\bm r}))\neq{\bm 0}$.
In other words, the detailed balance condition (potential condition~\cite{gardiner1985handbook}) is generally violated in this linear Langevin equation.
We can check that the detailed balance condition is satisfied when $\epsilon=0$.

\section{Quasi-cycles induced by stochastic amplification}
Although the coarse-grained deterministic description [Eqs.(\ref{rate eq_z}) and (\ref{rate eq_w})] cannot predict stable cycles, the stochastic description [Eqs.~(\ref{linear Langevin_z}) and (\ref{linear Langevin_w})] predicts persistent cyclic oscillations called \textit{quasi-cycles}.
Figure~\ref{fig:four_pictures}(c) shows an example of such persistent fluctuating oscillations observed in the linear Langevin equation.
Notably, the fluctuations of zonal flow follow those of turbulence with a phase lag $\sim\pi/2$, which are similar to the predator-prey-like oscillations observed in a toroidal plasma~\cite{estrada2010experimental,schmitz2012role,itoh2013assessment,manz2010long,manz2012zonal}.
Here, it should be noted that several experimental studies have also observed the opposite behavior, where turbulence follows zonal flow~\cite{conway2011mean,cheng2013dynamics,cavedon2016interplay}.
Whether such behavior can be reproduced in our model requires further investigation in future studies.

This resonant behavior can be quantified by the power spectral density matrix, which is defined as the Fourier transform of the steady-state autocorrelation function~\cite{gardiner1985handbook}:
\begin{align}
\mathsf{S}(\omega)&:=\dfrac{1}{2\pi}\int^\infty_{-\infty}\langle({\bm r}_\tau-\langle{\bm r}_\tau\rangle)({\bm r}_0-\langle{\bm r}_0\rangle)^\top\rangle e^{i\omega\tau}d\tau\notag\\
&=\dfrac{1}{2\pi}\Bigl(\mathsf{A}-i\omega\mathsf{I}\Bigr)^{-1}\mathsf{B}\mathsf{B}^\top\Bigl(\mathsf{A}^\top+i\omega\mathsf{I}\Bigr)^{-1}.
\end{align}
Then, the power spectral density for zonal flow $Z$ can be calculated as
\begin{align}
\mathsf{S}^{ZZ}(\omega)&=\dfrac{C_1+C_2\omega^2}{(\omega^2-\epsilon \beta)^2+(1-\epsilon)^2\omega^2},
\label{psd}
\end{align}
where
\begin{align}
C_1&:=\dfrac{1}{\pi}\Bigl((1-3\epsilon)D^Z+\epsilon^2 D^W\Bigr),\\
C_2&:=\dfrac{D^Z}{\pi}.
\end{align}
From this expression, we can see that the quasi-cycle occurs for a wide range of parameters where the characteristic resonant frequency $\omega_c:=\sqrt{\epsilon \beta-(1-\epsilon)^2/2}$ takes a real value (see also Fig.~\ref{fig:information_flow}(b)).
When this resonance condition is satisfied, the coexistence fixed point is a stable spiral fixed point with the characteristic frequency $|\mathrm{Im}[\lambda_\pm]|=\sqrt{\epsilon \beta-(1-\epsilon)^2/4}$, where $\mathrm{Im}[\cdot]$ denotes the imaginary part.
Note that no tuning is necessary to achieve resonance in this model.
Indeed, because the system is driven by the white noise, which covers all frequencies, the resonant frequency $\omega_c\sim|\mathrm{Im}[\lambda_\pm]|$ is excited without tuning.
This mechanism is called \textit{stochastic amplification} to distinguish it from \textit{stochastic resonance}~\cite{mckane2005predator}.
We remark that a similar analysis can be carried out for turbulence $W$.


Note that stochastic oscillations can also occur for a dynamical system with limit cycles perturbed by extrinsic noise. 
While quasi-cycles resemble such stochastic limit cycles, they possess several distinguishing characteristics.
First, the power spectral density of quasi-cycles exhibits a long tail with $\omega^{-2}$ scaling [Eq.~\ref{psd}]; in contrast, stochastic limit cycles decay more rapidly as $\omega^{-4}$~\cite{butler2011fluctuation}.
This implies that the correlation time scale of fluctuations in quasi-cycles is shorter than that of stochastic limit cycles~\cite{pineda2007tale}.
Second, the joint probability density of the oscillating variables for quasi-cycles is Gaussian [Eq.~(\ref{Gaussian})], whereas that for stochastic limit cycles takes the shape of a crater ridge~\cite{pineda2007tale}.
Furthermore, because quasi-cycles can readily emerge in any system possessing a stable spiral fixed point subject to noise, they may be more widely observed in plasma turbulence under various conditions.
Although we have emphasized that our model is a qualitative description, here we present a set of parameter values consistent with experiments and simulations for reference.
By comparing with the gyrokinetic simulations~\cite{kobayashi2015direct}, they can be estimated as $\epsilon\sim1$ and $\beta\sim10^{-4}$, which satisfies the resonance condition, and the corresponding cycle frequency reads $\omega_c\sim10^{-2}$ in dimensionless units.

\section{Information propagation between zonal flow and turbulence}
The presence of quasi-cycles suggests that intrinsic fluctuations are essential for understanding the dynamics of zonal flow and turbulence.
Therefore, we scrutinize the predictive causality underlying the intrinsic fluctuations from an information-theoretic viewpoint.
In the context of plasma turbulence, (net) transfer entropy~\cite{schreiber2000measuring,van2014causality} and information length~\cite{kim2020time,hollerbach2020time,kim2022non,kim2023stochastic,fuller2023time,fuller2024time,kim2024nonperturbative,kim2025probabilistic} have been used to quantify predictive causality.
Although these quantities are useful in quantifying various aspects of causal influences, it should be noted that they are not antisymmetric under time reversal.
In particular, the (net) transfer entropy can be nonzero even in an equilibrium state where all probability currents vanish~\cite{chetrite2019information}.
(For other shortcomings of the transfer entropy, see, e.g., Refs.~\cite{smirnov2013spurious,james2016information}.)
For the information length, although it becomes zero in such an equilibrium state, it also vanishes in a nonequilibrium steady state where some probability currents remain nonzero.
Therefore, from the perspective of consistency with the behavior of the probability currents, it is unclear whether these quantities appropriately quantify predictive causality.
In this study, we employ \textit{information flow}~\cite{allahverdyan2009thermodynamic,horowitz2014thermodynamics,parrondo2015thermodynamics,chetrite2019information} because it is antisymmetric under time reversal and becomes zero in the equilibrium state.

We first introduce the \textit{mutual information}~\cite{cover1999elements}, which quantifies the mutual dependence between zonal flow and turbulence:
\begin{align}
I[Z\colon\!W]:=\int dzdwp_\tau(z,w)\ln\dfrac{p_\tau(z,w)}{p^Z_\tau(z)p^W_\tau(w)}\ge0,
\end{align}
where $p_\tau(z,w)$ denotes the joint probability density, and $p^Z_\tau(z)$ and $p^W_\tau(w)$ denote the marginal distributions.
The mutual information is nonnegative and equal to zero if and only if zonal flow and turbulence are statistically independent. 

\begin{figure}[t]
\center
\includegraphics[width=8.6cm]{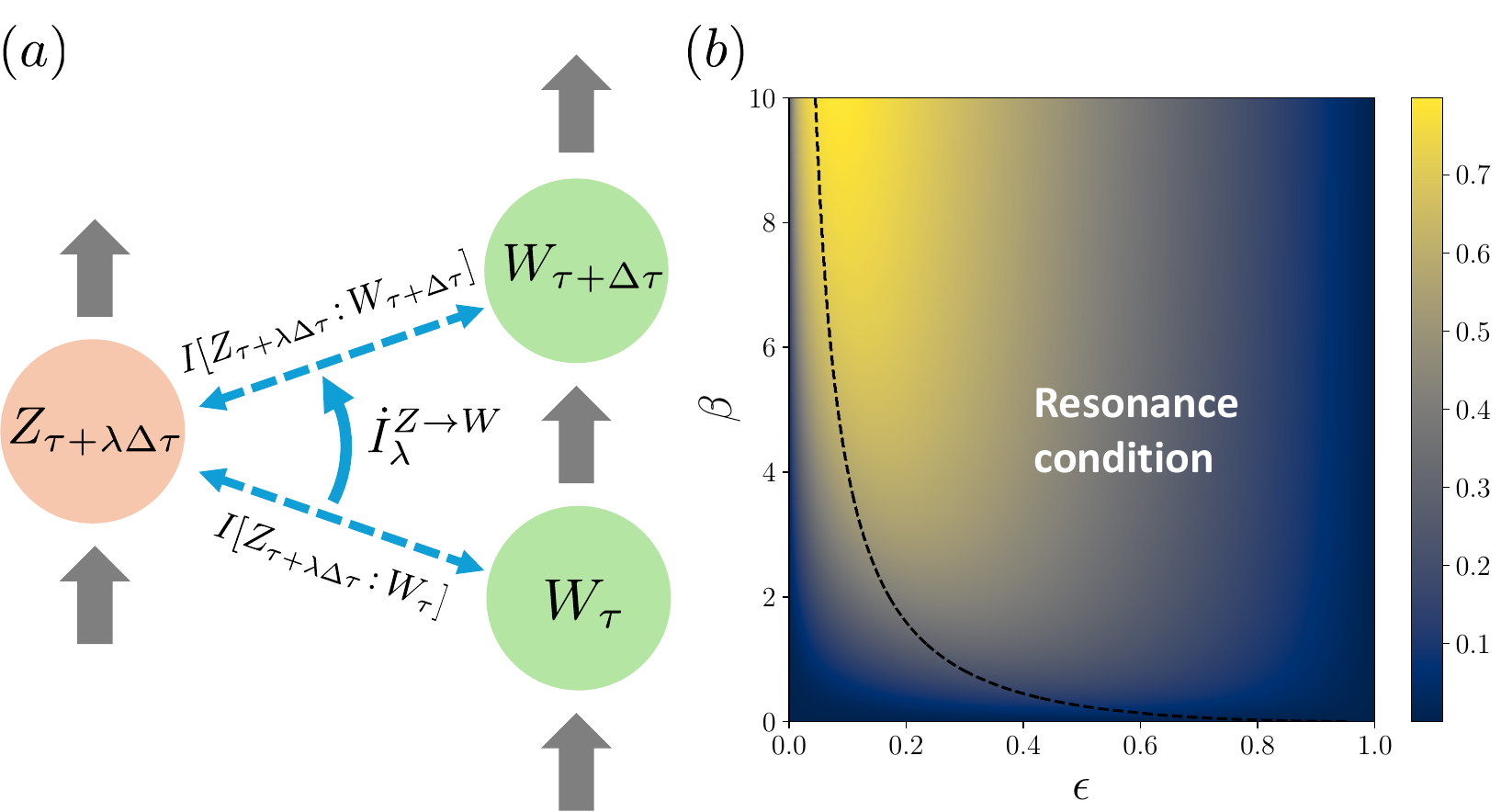}
\caption{(a) Schematic of information flow $\dot{I}^{Z\rightarrow W}_\lambda$.
(b) $(\epsilon,\beta)$-dependence of $\dot{I}^{Z\rightarrow W}_{\lambda=1/2}$ with $D^W=1/2$. The dashed line denotes the resonance condition $2\epsilon \beta-(1-\epsilon)^2=0$. The stochastic amplification occurs in the parameter region above this line.
}
\label{fig:information_flow}
\end{figure}

The decomposition of the time derivative of the mutual information yields information flow.
Specifically, the information flow from zonal flow to turbulence is defined as (see Fig.~\ref{fig:information_flow}(a))
\begin{align}
\dot{I}^{Z\rightarrow W}_\lambda:=\lim_{\Delta\tau\rightarrow0^+}\dfrac{I[Z_{\tau+\lambda\Delta\tau}\colon\!W_{\tau+\Delta\tau}]-I[Z_{\tau+\lambda\Delta\tau}\colon\!W_\tau]}{\Delta\tau}
\label{def IF}
\end{align}
with $\lambda\in[0,1]$.
We can similarly define the information flow from turbulence to zonal flow $\dot{I}^{W\rightarrow Z}_\lambda$, which is related to $\dot{I}^{Z\rightarrow W}_\lambda$ as $d_\tau I[Z\colon\!W]=\dot{I}^{Z\rightarrow W}_\lambda+\dot{I}^{W\rightarrow Z}_{1-\lambda}$.
Hence, we have $\dot{I}^{Z\rightarrow W}_\lambda=-\dot{I}^{W\rightarrow Z}_{1-\lambda}$ in the steady state.
Here, the information flow $\dot{I}^{Z\rightarrow W}_\lambda$ depends on $\lambda$ because the Langevin equations (\ref{linear Langevin_z}) and (\ref{linear Langevin_w}) do not satisfy the \textit{bipartite} condition.
That is, the noises in Eqs.~(\ref{linear Langevin_z}) and (\ref{linear Langevin_w}) are not independent, but correlated as $\langle\zeta^Z_\tau\zeta^W_{\tau'}\rangle=-D^Z\delta(\tau-\tau')$.
We remark that, in the context of information thermodynamics~\cite{parrondo2015thermodynamics,peliti2021stochastic,shiraishi2023introduction}, it is common to consider bipartite systems, where information flow does not depend on $\lambda$, and non-bipartite systems have rarely been studied~\cite{chetrite2019information}.

The information flow $\dot{I}^{Z\rightarrow W}_\lambda$ quantifies the rate at which turbulence gains information about zonal flow.
In particular, its sign indicates the direction in which the information is transmitted:
If $\dot{I}^{Z\rightarrow W}_\lambda>0$ for all $\lambda$, the information of zonal flow is transferred to turbulence, and vice versa.
Here, we note that the non-bipartite structure allows the case where the sign of $\dot{I}^{Z\rightarrow W}_\lambda$ varies depending on $\lambda$, which makes it difficult to determine the direction of information transfer~\cite{chetrite2019information}.
In that case, $\dot{I}^{Z\rightarrow W}_{\lambda=1/2}$ plays a special role because it is antisymmetric under time reversal and vanishes in the equilibrium state.
Fortunately, as will be shown below, this problem is irrelevant to our model.

From Eq.~(\ref{def IF}), we can express the information flow in terms of the probability current $J^W_{\tau}({\bm r})$ as
\begin{align}
\dot{I}^{Z\rightarrow W}_\lambda&=\int dzdw\left[J^W_{\tau}({\bm r})+(2\lambda-1)\dfrac{D^Z}{2}\dfrac{\partial}{\partial z}p_{\tau}({\bm r})\right]\notag\\
&\qquad\times\dfrac{\partial}{\partial w}\ln\dfrac{p_{\tau}({\bm r})}{p_{\tau}(z)p_{\tau}(w)},\label{lambda-IF_W}
\end{align}
where $p_{\tau}(z)$ and $p_{\tau}(w)$ are marginal distributions for zonal flow and turbulence, respectively.
By noting that the steady-state distribution $p_{\mathrm{ss}}({\bm r})$ is Gaussian [Eq.~(\ref{Gaussian})], the information flow in the steady state can be calculated as
\begin{align}
\dot{I}^{Z\rightarrow W}_\lambda&=\dfrac{D^Z}{\det\Sigma}\left[\Sigma^{WW}+(1-\lambda)\Sigma^{ZW}\right]\notag\\
&>0\quad\text{for all}\quad\lambda\in[0,1],
\label{main result}
\end{align}
where $\Sigma^{ZW}=\langle(z-\langle z\rangle)(w-\langle w\rangle)\rangle$ and $\Sigma^{WW}=\langle(w-\langle w\rangle)^2\rangle$.
See Appendix~\ref{Detailed calculation of information flow} for the detailed derivation of Eqs.~(\ref{lambda-IF_W}) and (\ref{main result}).
The inequality (\ref{main result}) states that the information about fluctuations of zonal flow is propagated to turbulence.
In other words, the turbulence is ``learning'' about the fluctuating behavior of the zonal flow.
This result is somewhat counterintuitive because the system exhibits persistent predator-prey cycles, where the fluctuations of zonal flow follow those of turbulence with a time lag.
Figure~\ref{fig:information_flow}(b) shows the $(\epsilon,\beta)$-dependence of $\dot{I}^{Z\rightarrow W}_{\lambda=1/2}$.
Interestingly, Fig.~\ref{fig:information_flow}(b) suggests that the behavior of the information flow does not depend significantly on whether stochastic amplification occurs or not.
In other words, even when the system does not exhibit the fluctuating predator-prey oscillation, there is a predictive causality from zonal flow to turbulence.
This information-theoretic analysis may provide a theoretical foundation for the regulation of turbulence through the control of zonal flow.
We finally remark that the inferred causality from the information flow is consistent with that from the cross-correlation function, although interpreting the latter is less straightforward (see Appendix~\ref{Comparison with cross-correlation function}).

\section{Concluding remarks}
We have proposed a simple stochastic predator-prey model for plasma turbulence that exhibits quasi-cycles induced by stochastic amplification of intrinsic noise.
For this model, we have quantified the predictive causality between zonal flow and turbulence by showing that the information of zonal flow is propagated to turbulence.

Several remarks on our results are in order.
First, although we have considered the four simple ``reactions'' that incorporate a minimal self-regulation mechanism, it is straightforward to extend our approach to more complicated reactions, such as those involving time-dependent rate constants or additional reactant species (e.g., the mean shear flow).
Even in such a case, both the system size expansion and the information-theoretic analysis are applicable.

Second, our model illustrates the significance of considering the effects of intrinsic noise in understanding the predator-prey-like oscillations observed in plasma turbulence.
Notably, our results suggest the possibility that the observed cyclic oscillations in a toroidal plasma are quasi-cycles rather than limit cycles.
Using the methods proposed in Ref.~\cite{pineda2007tale}, it would be possible to distinguish quasi-cycles and (stochastic) limit cycles both in experiments and simulations.
Further studies are needed to verify whether both quasi-cycles and information transfer from zonal flow to turbulence can be observed in experiments and simulations of more realistic models, such as the Hasegawa--Wakatani equation~\cite{hasegawa1983plasma}.

Third, our results essentially hold true for general predator-prey dynamics, not limited to plasma turbulence.
In particular, while the essence of our first main result---the emergence of quasi-cycles due to stochastic amplification in the predator-prey dynamics---has already been recognized in the field of ecology~\cite{mckane2005predator,black2012stochastic}, our second main result concerning information flow may offer new perspectives even within an ecological context.

Finally, while we have shown that information of zonal flow is propagated to turbulence, further research is needed to understand how this result relates to the regulation of turbulence through the control of zonal flow.
This issue also seems connected to the general question of how predictive causality relates to dynamical causality, which can be quantified in frameworks such as convergent cross mapping~\cite{sugihara2012detecting,dumerat2025causal}.
We hope our research will serve as a starting point for investigating these issues.

\begin{acknowledgments}
We thank S.~Maeyama and M.~Nakata for fruitful discussions.
T.T.~was supported by JSPS KAKENHI Grant Number JP25K17315 and JST PRESTO Grant Number JPMJPR23O6, Japan.
M.S.~was supported by JSPS KAKENHI Grant Number JP25K00986, Japan.
\end{acknowledgments}

\appendix
\section{Detailed calculation of system size expansion\label{Detailed calculation of system size expansion}}
We consider the macroscopic limit of the large system size $\Omega$.
In this limit, the probability $p_t({\bm n})$ can be replaced with the probability density $p_t(\tilde{\bm c})=\Omega^2p_t({\bm n})$, where $\tilde{\bm c}=(\tilde{c}^Z,\tilde{c}^W):={\bm n}/\Omega\in\mathbb{R}^2_{\ge0}$ denotes a continuous variable representing the energy densities of $Z$ and $W$.
Correspondingly, we introduce the intensive transition rate $\mathcal{W}_\rho\left(\tilde{\bm c}\right)$ as
\begin{align}
\mathcal{W}_\rho\left(\tilde{\bm c}\right)&:=\lim_{\Omega\rightarrow\infty}\dfrac{W_\rho({\bm n}+{\bm S}_\rho|{\bm n})}{\Omega}\notag\\
&=\rho(\tilde{c}^Z)^{\nu^Z_\rho}(\tilde{c}^W)^{\nu^W_\rho}.
\end{align}
Then, by taking the Kramers--Moyal expansion~\cite{gardiner1985handbook}, the master equation [Eq.~(\ref{master eq})] can be rewritten as
\begin{align}
\partial_tp_t(\tilde{\bm c})=\sum_\rho\sum^\infty_{k=1}\dfrac{\Omega}{k!}\left(-\dfrac{{\bm S}_\rho}{\Omega}\cdot\dfrac{\partial}{\partial \tilde{\bm c}}\right)^k\mathcal{W}_\rho (\tilde{\bm c})p_t(\tilde{\bm c}).
\label{Kramers-Moyal expansion}
\end{align}

From the large deviation principle, we expect that $p_t(\tilde{\bm c})$ acquires the following form in the macroscopic limit $\Omega\rightarrow\infty$:
\begin{align}
p_t(\tilde{\bm c})\asymp e^{-\Omega I_t(\tilde{\bm c})},
\end{align}
where the rate function $I_t(\tilde{\bm c})\ge0$ is $\Omega$-independent, and the symbol $\asymp$ denotes the logarithm equality, i.e., $-\lim_{\Omega\rightarrow\infty}\Omega^{-1}\ln p_t(\tilde{\bm c})=I_t(\tilde{\bm c})$~\cite{falasco2023macroscopic}.
In other words, $p_t(\tilde{\bm c})$ will be a $\delta$ function in this limit, where the system is described by the deterministic dynamics of the minima of $I_t(\tilde{\bm c})$.
For large, but finite $\Omega$, $p_t(\tilde{\bm c})$ may have a finite width of order $\Omega^{-1/2}$.
To describe the fluctuating dynamics around the typical value of $\tilde{\bm c}$, we employ van Kampen's system size expansion~\cite{van1992stochastic,gardiner1985handbook}.
We first introduce a new variable ${\bm r}=(z,w)\in\mathbb{R}^2$, which can be interpreted as the fluctuations in the energy densities, such that
\begin{align}
\tilde{\bm c}={\bm c}+\dfrac{\bm r}{\Omega^{1/2}},
\end{align}
where the mean-field density ${\bm c}\in\mathbb{R}^2_{\ge0}$ is a function of time $t$ to be determined.
We denote by ${\bm c}_t$ and ${\bm r}_t$ the value of ${\bm c}$ and ${\bm r}$ at time $t$, respectively.
Correspondingly, let $p_t({\bm r}):=\Omega^{-1/2}p_t({\bm c}_t+\Omega^{-1/2}{\bm r})$ be the probability density of finding ${\bm r}_t={\bm r}$.
Then, Eq.~(\ref{Kramers-Moyal expansion}) can be expressed as
\begin{align}
&\partial_tp_t({\bm r})-\Omega^{1/2}\dot{\bm c}_t\cdot\dfrac{\partial}{\partial {\bm r}}p_t({\bm r})\notag\\
&=\sum_\rho\sum^\infty_{k=1}\dfrac{\Omega^{1-k/2}}{k!}\left(-{\bm S}_\rho\cdot\dfrac{\partial}{\partial {\bm r}}\right)^k\mathcal{W}_\rho\left({\bm c}_t+\dfrac{\bm r}{\Omega^{1/2}}\right)p_t({\bm r}).
\label{ME_ss_expansion}
\end{align}

\subsection{Deterministic rate equation}
The leading order $O(\Omega^{1/2})$ of Eq.~(\ref{ME_ss_expansion}) yields the deterministic rate equation for the average part ${\bm c}_t=(c^Z_t,c^W_t)$:
\begin{align}
\dfrac{d}{dt}{\bm c}_t=\sum_\rho{\bm S}_\rho \mathcal{W}_\rho({\bm c}_t).
\end{align}
By noting that $\mathcal{W}_{d}({\bm c})=dc^Z,\mathcal{W}_b({\bm c})=bc^W,\mathcal{W}_p({\bm c})=pc^Zc^W$, and $\mathcal{W}_\gamma({\bm c})=\gamma (c^W)^2$, the rate equation can be expressed as
\begin{align}
\dfrac{d}{dt}c^Z_t&=p c^Z_tc^W_t-d c^Z_t,\label{appendix: rate eq_z}\\
\dfrac{d}{dt}c^W_t&=b c^W_t-\gamma (c^W_t)^2-p c^Z_tc^W_t,\label{appendix: rate eq_w}
\end{align}
which corresponds to Eqs.~(\ref{rate eq_z}) and (\ref{rate eq_w}).

\subsection{Linear Langevin equation}
The subleading order $O(1)$ of Eq.~(\ref{ME_ss_expansion}) yields the Fokker--Planck equation:
\begin{align}
&\partial_tp_t({\bm r})=-\sum_\rho\left.\dfrac{\partial}{\partial {\bm c}}\mathcal{W}_\rho({\bm c})\right|_{{\bm c}={\bm c}_t}\cdot\left({\bm S}_\rho\cdot\dfrac{\partial}{\partial {\bm r}}\right){\bm r}p_t({\bm r})\notag\\
&\qquad\qquad\quad+\sum_\rho\dfrac{1}{2}\mathcal{W}_\rho({\bm c}_t)\left({\bm S}_\rho\cdot\dfrac{\partial}{\partial {\bm r}}\right)^2p_t({\bm r}).
\label{subleading order ME_ss_expansion}
\end{align}
Note that ${\bm c}_t$ in Eq.~(\ref{subleading order ME_ss_expansion}) obeys the deterministic rate equations (\ref{appendix: rate eq_z}) and (\ref{appendix: rate eq_w}).
In other words, the nonlinearity of the predator-prey dynamics is reflected in the Fokker--Planck equation through the dependence on ${\bm c}_t$. 
If we substitute the fixed point solution ${\bm c}_t={\bm c}_{\mathrm{ss}}$, then the Fokker--Planck equation (\ref{subleading order ME_ss_expansion}) describes the time evolution of the fluctuations around the fixed point.
Since we are interested in the state where zonal flow and turbulence coexist, we focus on the regime $bp>d\gamma$ and substitute the stable coexistence fixed point $(c^Z_{\mathrm{ss}},c^W_{\mathrm{ss}})=((bp-d\gamma)/p^2,d/p)$ into Eq.~(\ref{subleading order ME_ss_expansion}).
By introducing the dimensionless time $\tau:=bt$ and dimensionless parameters $\beta:=d/b>0$ and $\epsilon:=1-d\gamma/bp$ ($0<\epsilon<1$), the resulting Fokker--Planck equation can be expressed as
\begin{align}
\partial_\tau p_\tau({\bm r})=-\dfrac{\partial}{\partial z}J^Z_\tau({\bm r})-\dfrac{\partial}{\partial w}J^W_\tau({\bm r}),
\label{FP}
\end{align}
where $J^Z_\tau({\bm r})$ and $J^W_\tau({\bm r})$ denote the dimensionless probability currents defined by
\begin{align}
J^Z_\tau({\bm r})&:=\epsilon wp_\tau({\bm r})-D^Z\dfrac{\partial}{\partial z}p_\tau({\bm r})+\dfrac{D^Z}{2}\dfrac{\partial}{\partial w}p_\tau({\bm r}),\label{dimensionless_probability current_Z}\\
J^W_\tau({\bm r})&:=\bigl[-\beta z-(1-\epsilon)w\bigr]p_\tau({\bm r})\notag\\
&\qquad-D^W\dfrac{\partial}{\partial w}p_\tau({\bm r})+\dfrac{D^Z}{2}\dfrac{\partial}{\partial z}p_\tau({\bm r}),
\end{align}
with $D^Z:=\epsilon c^W_{\mathrm{ss}}=\epsilon d/p$ and $D^W:=c^W_{\mathrm{ss}}=d/p$.
The Langevin equation corresponding to this Fokker--Planck equation reads
\begin{align}
\dfrac{d}{d\tau}z_\tau&=\epsilon w_\tau+\zeta^Z_\tau,\\
\dfrac{d}{d\tau}w_\tau&=-\beta z_\tau-(1-\epsilon)w_\tau+\zeta^W_\tau,
\end{align}
where $\zeta^Z_\tau$ and $\zeta^W_\tau$ denote the zero-mean white Gaussian noise that satisfies
\begin{align}
\langle\zeta^Z_\tau\zeta^Z_{\tau'}\rangle&=2D^Z\delta(\tau-\tau'),\\
\langle\zeta^Z_\tau\zeta^W_{\tau'}\rangle&=-D^Z\delta(\tau-\tau'),\\
\langle\zeta^W_\tau\zeta^W_{\tau'}\rangle&=2D^W\delta(\tau-\tau').
\end{align}
Note that $\beta$ and $\epsilon$ can be interpreted as a dimensionless damping rate and the ratio of time scales for $z$ and $w$, respectively.
Here, we emphasize that the noise intensities $D^Z$ and $D^W$ are determined from the rate constants $\rho$ in the master equation [Eq.~(\ref{master eq})].
In other words, the noise that drives the system is not external but arises from the intrinsic stochasticity.

We finally remark that the validity of the system size expansion up to the subleading order has been well confirmed in the literature.
For example, in Ref.~\cite{wallace2012linear}, it is reported that the linear Langevin equation agrees well with the master equation for sufficiently large system size.

\begin{widetext}
\section{Detailed calculation of information flow\label{Detailed calculation of information flow}}
\subsection{Derivation of Eq.~(\ref{lambda-IF_W})}
In this section, we derive Eq.~(\ref{lambda-IF_W}). 
From the definition of $\dot{I}^{Z\rightarrow W}_\lambda$, we first note that
\begin{align}
\dot{I}^{Z\rightarrow W}_\lambda&:=\lim_{\Delta\tau\rightarrow0^+}\dfrac{I[Z_{\tau+\lambda\Delta\tau}\colon\!W_{\tau+\Delta\tau}]-I[Z_{\tau+\lambda\Delta\tau}\colon\!W_\tau]}{\Delta\tau}\notag\\
&=\lim_{h\rightarrow0^+}\dfrac{1}{h}\left(\int dzdwp(z,\tau+\lambda h;w,\tau+h)\ln\dfrac{p(z,\tau+\lambda h;w,\tau+h)}{p_{\tau+\lambda h}(z)p_{\tau+h}(w)}-\int dzdwp(z,\tau+\lambda h;w,\tau)\ln\dfrac{p(z,\tau+\lambda h;w,\tau)}{p_{\tau+\lambda h}(z)p_\tau(w)}\right),
\end{align}
where $p(z,\tau+\lambda h;w,\tau+h)$ and $p(z,\tau+\lambda h;w,\tau)$ denote the two-point probability densities.
When $h=0$, the two-point probability densities correspond to the joint probability density at time $\tau$: $p(z,\tau;w,\tau)=p_\tau(z,w)$.
By expanding $p(z,\tau+\lambda h;w,\tau+h)$, $p(z,\tau+\lambda h;w,\tau)$, $p_{\tau+h}(w)$, and $p_{\tau+\lambda h}(z)$ with respect to $h$, we obtain
\begin{align}
\dot{I}^{Z\rightarrow W}_\lambda&=\int dz'dw\left.\dfrac{d}{dh}p(z',\tau+\lambda h;w,\tau+h)\right|_{h=0}\ln\dfrac{p_\tau(z',w)}{p_\tau(z')p_\tau(w)}\notag\\
&\qquad-\int dzdw'\left.\dfrac{d}{dh}p(z,\tau+\lambda h;w',\tau)\right|_{h=0}\ln\dfrac{p_\tau(z,w')}{p_\tau(z)p_\tau(w')}.
\label{IF expanding in h}
\end{align}
Here, for later convenience, we have replaced $z$ and $w$ with $z'$ and $w'$ in the first and second integrals, respectively.
By noting that
\begin{align}
p(z',\tau+\lambda h;w,\tau+h)-p(z',\tau;w,\tau)&=p(z',\tau+\lambda h;w,\tau+\lambda h+(1-\lambda)h)-p(z',\tau+\lambda h;w,\tau+\lambda h)\notag\\
&\qquad+p(z',\tau+\lambda h;w,\tau+\lambda h)-p(z',\tau;w,\tau),
\end{align}
the derivative $\frac{d}{dh}p(z',\tau+\lambda h;w,\tau+h)|_{h=0}$ can be expressed as
\begin{align}
\left.\dfrac{d}{dh}p(z',\tau+\lambda h;w,\tau+h)\right|_{h=0}=(1-\lambda)\left.\dfrac{d}{dh}p(z',\tau;w,\tau+h)\right|_{h=0}+\lambda\left.\dfrac{d}{dh}p_{\tau+h}(z',w)\right|_{h=0}.
\end{align}
By substituting this expression into Eq.~(\ref{IF expanding in h}), we have
\begin{align}
\dot{I}^{Z\rightarrow W}_\lambda&=(1-\lambda)\int dz'dw\left.\dfrac{d}{dh}p(z',\tau;w,\tau+h)\right|_{h=0}\ln\dfrac{p_\tau(z',w)}{p_\tau(z')p_\tau(w)}\notag\\
&\qquad+\lambda\int dz'dw\left.\dfrac{d}{dh}p_{\tau+h}(z',w)\right|_{h=0}\ln\dfrac{p_\tau(z',w)}{p_\tau(z')p_\tau(w)}\notag\\
&\qquad-\lambda\int dzdw'\left.\dfrac{d}{dh}p(z,\tau+h;w',\tau)\right|_{h=0}\ln\dfrac{p_\tau(z,w')}{p_\tau(z)p_\tau(w')}\notag\\
&=(1-\lambda)\int dz'dw'dzdw\left.\dfrac{d}{dh}p(z,w,\tau+h|z',w',\tau)\right|_{h=0}p_\tau(z',w')\ln\dfrac{p_\tau(z',w)}{p_\tau(z')p_\tau(w)}\notag\\
&\qquad+\lambda\int dzdw\left.\dfrac{d}{dh}p_{\tau+h}(z,w)\right|_{h=0}\ln\dfrac{p_\tau(z,w)}{p_\tau(z)p_\tau(w)}\notag\\
&\qquad-\lambda\int dz'dw'dzdw\left.\dfrac{d}{dh}p(z,w,\tau+h|z',w',\tau)\right|_{h=0}p_\tau(z',w')\ln\dfrac{p_\tau(z,w')}{p_\tau(z)p_\tau(w')}.
\label{IF expanding in h_2}
\end{align}
In the second equality, we have used
\begin{align}
p(z',\tau;w,\tau+h)&=\int dw'dzp(z',w',\tau;z,w,\tau+h)=\int dw'dzp(z,w,\tau+h|z',w',\tau)p_\tau(z',w'),\\
p(z,\tau+h;w',\tau)&=\int dwdz'p(z,w,\tau+h;z',w',\tau)=\int dwdz'p(z,w,\tau+h|z',w',\tau)p_\tau(z',w'),
\end{align}
where $p(z,w,\tau+h|z',w',\tau)$ denotes the conditional probability density.
Note that this conditional probability density obeys the Fokker--Planck equation~(\ref{FP}):
\begin{align}
\left.\dfrac{d}{dh}p(z,w,\tau+h|z',w',\tau)\right|_{h=0}&=-\partial_z\left[F^Z({\bm r})\delta({\bm r}-{\bm r}')-D^Z\partial_z\delta({\bm r}-{\bm r}')+\dfrac{D^Z}{2}\partial_w\delta({\bm r}-{\bm r}')\right]\notag\\
&\qquad-\partial_w\left[F^W({\bm r})\delta({\bm r}-{\bm r}')-D^W\partial_w\delta({\bm r}-{\bm r}')+\dfrac{D^Z}{2}\partial_z\delta({\bm r}-{\bm r}')\right],
\end{align}
where $F^Z({\bm r}):=\epsilon w$ and $F^W({\bm r}):=-\beta z-(1-\epsilon)w$, and we have used $p(z,w,\tau|z',w',\tau)=\delta({\bm r}-{\bm r}')$.
Then, Eq.~(\ref{IF expanding in h_2}) can be calculated as
\begin{align}
\dot{I}^{Z\rightarrow W}_\lambda&=-(1-\lambda)\int dz'dw'dzdw\partial_z\left[F^Z({\bm r})\delta({\bm r}-{\bm r}')-D^Z\partial_z\delta({\bm r}-{\bm r}')+\dfrac{D^Z}{2}\partial_w\delta({\bm r}-{\bm r}')\right]p_\tau(z',w')\ln\dfrac{p_\tau(z',w)}{p_\tau(z')p_\tau(w)}\notag\\
&\quad-(1-\lambda)\int dz'dw'dzdw\partial_w\left[F^W({\bm r})\delta({\bm r}-{\bm r}')-D^W\partial_w\delta({\bm r}-{\bm r}')+\dfrac{D^Z}{2}\partial_z\delta({\bm r}-{\bm r}')\right]p_\tau(z',w')\ln\dfrac{p_\tau(z',w)}{p_\tau(z')p_\tau(w)}\notag\\
&\quad+\lambda\int dzdw\left[-\partial_zJ^Z_\tau({\bm r})-\partial_wJ^W_\tau({\bm r})\right]\ln\dfrac{p_\tau(z,w)}{p_\tau(z)p_\tau(w)}\notag\\
&\quad+\lambda\int dz'dw'dzdw\partial_z\left[F^Z({\bm r})\delta({\bm r}-{\bm r}')-D^Z\partial_z\delta({\bm r}-{\bm r}')+\dfrac{D^Z}{2}\partial_w\delta({\bm r}-{\bm r}')\right]p_\tau(z',w')\ln\dfrac{p_\tau(z,w')}{p_\tau(z)p_\tau(w')}\notag\\
&\quad+\lambda\int dz'dw'dzdw\partial_w\left[F^W({\bm r})\delta({\bm r}-{\bm r}')-D^W\partial_w\delta({\bm r}-{\bm r}')+\dfrac{D^Z}{2}\partial_z\delta({\bm r}-{\bm r}')\right]p_\tau(z',w')\ln\dfrac{p_\tau(z,w')}{p_\tau(z)p_\tau(w')}\notag\\
&=-(1-\lambda)\int dzdw\partial_z\left[F^Z({\bm r})p_\tau(z,w)\ln\dfrac{p_\tau(z,w)}{p_\tau(z)p_\tau(w)}-D^Z\partial_z\left(p_\tau(z,w)\ln\dfrac{p_\tau(z,w)}{p_\tau(z)p_\tau(w)}\right)\right.\notag\\
&\qquad\left.+\dfrac{D^Z}{2}\biggl(\partial_wp_\tau(z,w)\biggr)\ln\dfrac{p_\tau(z,w)}{p_\tau(z)p_\tau(w)}\right]\notag\\
&\quad-(1-\lambda)\int dzdw\left\{\partial_w\left[F^W({\bm r})p_\tau(z,w)-D^W\partial_wp_\tau(z,w)\right]\ln\dfrac{p_\tau(z,w)}{p_\tau(z)p_\tau(w)}+\dfrac{D^Z}{2}\partial_z\left[\biggl(\partial_wp_\tau(z,w)\biggr)\ln\dfrac{p_\tau(z,w)}{p_\tau(z)p_\tau(w)}\right]\right\}\notag\\
&\quad+\lambda\int dzdw\left[-\partial_zJ^Z_\tau({\bm r})-\partial_wJ^W_\tau({\bm r})\right]\ln\dfrac{p_\tau(z,w)}{p_\tau(z)p_\tau(w)}\notag\\
&\quad+\lambda\int dzdw\left\{\partial_z\left[F^Z({\bm r})p_\tau(z,w)-D^Z\partial_zp_\tau(z,w)\right]\ln\dfrac{p_\tau(z,w)}{p_\tau(z)p_\tau(w)}+\dfrac{D^Z}{2}\partial_w\left[\biggl(\partial_zp_\tau(z,w)\biggr)\ln\dfrac{p_\tau(z,w)}{p_\tau(z)p_\tau(w)}\right]\right\}\notag\\
&\quad+\lambda\int dzdw\partial_w\left[F^W({\bm r})p_\tau(z,w)\ln\dfrac{p_\tau(z,w)}{p_\tau(z)p_\tau(w)}-D^W\partial_w\left(p_\tau(z,w)\ln\dfrac{p_\tau(z,w)}{p_\tau(z)p_\tau(w)}\right)\right.\notag\\
&\qquad\left.+\dfrac{D^Z}{2}\biggl(\partial_zp_\tau(z,w)\biggr)\ln\dfrac{p_\tau(z,w)}{p_\tau(z)p_\tau(w)}\right].
\end{align}
Note that, in the second equality, the first and last integrals and the last terms on both the second and fourth integrals vanish, assuming the natural boundary condition.
Then, we have
\begin{align}
\dot{I}^{Z\rightarrow W}_\lambda&=-(1-\lambda)\int dzdw\partial_w\left[F^W({\bm r})p_\tau(z,w)-D^W\partial_wp_\tau(z,w)\right]\ln\dfrac{p_\tau(z,w)}{p_\tau(z)p_\tau(w)}\notag\\
&\quad+\lambda\int dzdw\left[-\partial_zJ^Z_\tau({\bm r})-\partial_wJ^W_\tau({\bm r})\right]\ln\dfrac{p_\tau(z,w)}{p_\tau(z)p_\tau(w)}\notag\\
&\quad+\lambda\int dzdw\partial_z\left[F^Z({\bm r})p_\tau(z,w)-D^Z\partial_zp_\tau(z,w)\right]\ln\dfrac{p_\tau(z,w)}{p_\tau(z)p_\tau(w)}.
\end{align}
By noting that
\begin{align}
J^Z_\tau({\bm r})&:=F^Z({\bm r})p_\tau({\bm r})-D^Z\partial_zp_\tau({\bm r})+\dfrac{D^Z}{2}\partial_wp_\tau({\bm r}),\label{dimensionless_probability current_Z}\\
J^W_\tau({\bm r})&:=F^W({\bm r})p_\tau({\bm r})-D^W\partial_wp_\tau({\bm r})+\dfrac{D^Z}{2}\partial_zp_\tau({\bm r}),
\end{align}
we obtain
\begin{align}
\dot{I}^{Z\rightarrow W}_\lambda&=-\int dzdw\partial_w\left[F^W({\bm r})p_\tau(z,w)-D^W\partial_wp_\tau(z,w)+\lambda D^Z\partial_zp_\tau(z,w)\right]\ln\dfrac{p_\tau(z,w)}{p_\tau(z)p_\tau(w)}\notag\\
&=\int dzdw\left[J^W_{\tau}({\bm r})+(2\lambda-1)\dfrac{D^Z}{2}\partial_zp_{\tau}({\bm r})\right]\partial_w\ln\dfrac{p_{\tau}({\bm r})}{p_{\tau}(z)p_{\tau}(w)}.
\label{IF_lambda_W}
\end{align}
We thus arrive at Eq.~(\ref{lambda-IF_W}).
Similarly, we can also derive the following expression for $\dot{I}^{W\rightarrow Z}_{1-\lambda}$:
\begin{align}
\dot{I}^{W\rightarrow Z}_{1-\lambda}=\int dzdw\left[J^Z_{\tau}({\bm r})-(2\lambda-1)\dfrac{D^Z}{2}\partial_wp_{\tau}({\bm r})\right]\partial_z\ln\dfrac{p_{\tau}({\bm r})}{p_{\tau}(z)p_{\tau}(w)}.
\label{IF_lambda_Z}
\end{align}
We can easily check that Eqs.~(\ref{IF_lambda_W}) and (\ref{IF_lambda_Z}) satisfy $d_\tau I[Z\colon\!W]=\dot{I}^{Z\rightarrow W}_\lambda+\dot{I}^{W\rightarrow Z}_{1-\lambda}$.
Finally, we remark that $\dot{I}^{Z\rightarrow W}_\lambda$ can also be calculated from $\dot{I}^{Z\rightarrow W}_{\lambda=0}$ and $\dot{I}^{Z\rightarrow W}_{\lambda=1}$ as follows~\cite{chetrite2019information}:
\begin{align}
\dot{I}^{Z\rightarrow W}_\lambda=(1-\lambda)\dot{I}^{Z\rightarrow W}_{\lambda=0}+\lambda\dot{I}^{Z\rightarrow W}_{\lambda=1}.
\end{align}

\subsection{Derivation of Eq.~(\ref{main result})}
In this section, we derive Eq.~(\ref{main result}). 
The stationary joint distribution $p_{\mathrm{ss}}({\bm r})$ and marginal distributions $p_{\mathrm{ss}}(z)$ and $p_{\mathrm{ss}}(w)$ can be explicitly calculated as~\cite{gardiner1985handbook}
\begin{align}
p_{\mathrm{ss}}({\bm r})&=\dfrac{1}{2\pi\sqrt{\det\Sigma}}\exp\left(-\dfrac{1}{2}({\bm r}-\langle{\bm r}\rangle)^\top\Sigma^{-1}({\bm r}-\langle{\bm r}\rangle)\right),\label{joint distribution_ss}\\
p_{\mathrm{ss}}(z)&=\dfrac{1}{\sqrt{2\pi\Sigma^{ZZ}}}\exp\left(-\dfrac{1}{2}\dfrac{(z-\langle z\rangle)^2}{\Sigma^{ZZ}}\right),\label{marginal distribution_ss_Z}\\
p_{\mathrm{ss}}(w)&=\dfrac{1}{\sqrt{2\pi\Sigma^{WW}}}\exp\left(-\dfrac{1}{2}\dfrac{(w-\langle w\rangle)^2}{\Sigma^{WW}}\right),\label{marginal distribution_ss_W}
\end{align}
where $\Sigma$ denotes the covariance matrix
\begin{align}
\Sigma=
\begin{pmatrix}
\Sigma^{ZZ} & \Sigma^{ZW} \\
\Sigma^{WZ} & \Sigma^{WW}
\end{pmatrix}
=
\begin{pmatrix}
\langle(z-\langle z\rangle)^2\rangle & \langle(z-\langle z\rangle)(w-\langle w\rangle)\rangle \\
\langle(z-\langle z\rangle)(w-\langle w\rangle)\rangle & \langle(w-\langle w\rangle)^2\rangle
\end{pmatrix},
\end{align}
which satisfies the Lyapunov equation:
\begin{align}
\mathsf{A}\Sigma+\Sigma\mathsf{A}^\top=\mathsf{B}\mathsf{B}^\top,
\label{Lyapunov equation}
\end{align}
where $\mathsf{A}$ and $\mathsf{B}$ are matrices defined by Eqs.~(\ref{def A}) and (\ref{def B}).
By solving this equation, the covariance matrix can be calculated as~\cite{gardiner1985handbook}
\begin{align}
\Sigma&=\dfrac{(\det\mathsf{A})\mathsf{B}\mathsf{B}^\top+(\mathsf{A}-(\mathop{\mathrm{tr}}\mathsf{A})\mathsf{I})\mathsf{B}\mathsf{B}^\top(\mathsf{A}-(\mathop{\mathrm{tr}}\mathsf{A})\mathsf{I})^\top}{2(\mathop{\mathrm{tr}}\mathsf{A})(\det\mathsf{A})}\notag\\
&=\dfrac{1}{2\beta\epsilon(1-\epsilon)}
\begin{pmatrix}
2D^Z\left[\beta\epsilon+(1-\epsilon)(2(1-\epsilon)-1)\right]+2D^W\epsilon^2 & -2\beta(1-\epsilon)D^Z \\
-2\beta(1-\epsilon)D^Z & 2D^W\beta\epsilon+2\beta^2D^Z
\end{pmatrix}.
\label{Covariance matrix}
\end{align}

To derive Eq.~(\ref{main result}), it is easy to calculate $\dot{I}^{W\rightarrow Z}_{1-\lambda}$ instead of $\dot{I}^{Z\rightarrow W}_\lambda$ by using the relation $\dot{I}^{W\rightarrow Z}_{1-\lambda}=-\dot{I}^{Z\rightarrow W}_\lambda$ in the steady state.
By substituting (\ref{dimensionless_probability current_Z}), (\ref{joint distribution_ss}), and (\ref{marginal distribution_ss_Z}) into (\ref{IF_lambda_Z}), we can explicitly calculate the information flow as follows:
\begin{align}
\dot{I}^{Z\rightarrow W}_\lambda=-\dot{I}^{W\rightarrow Z}_{1-\lambda}&=-\int dzdw\left(\epsilon wp_\mathrm{ss}({\bm r})-D^Z\partial_zp_\mathrm{ss}({\bm r})+(1-\lambda)D^Z\partial_wp_\mathrm{ss}({\bm r})\right)\notag\\
&\qquad\times\left(-\dfrac{\Sigma^{WW}}{\det\Sigma}(z-\langle z\rangle)+\dfrac{\Sigma^{ZW}}{\det\Sigma}(w-\langle w\rangle)+\dfrac{z-\langle z\rangle}{\Sigma^{ZZ}}\right)\notag\\
&=-\epsilon\left[-\dfrac{\Sigma^{WW}\Sigma^{ZW}}{\det\Sigma}+\dfrac{\Sigma^{WW}\Sigma^{ZW}}{\det\Sigma}+\dfrac{\Sigma^{ZW}}{\Sigma^{ZZ}}\right]-D^Z\left[-\dfrac{\Sigma^{WW}}{\det\Sigma}+\dfrac{1}{\Sigma^{ZZ}}\right]+(1-\lambda)D^Z\dfrac{\Sigma^{ZW}}{\det\Sigma}\notag\\
&=\dfrac{D^Z}{\det\Sigma}\left[\Sigma^{WW}+(1-\lambda)\Sigma^{ZW}\right],
\end{align}
where we used $\Sigma^{ZW}=-D^Z/\epsilon$, which follows from (\ref{Covariance matrix}), in the last line.
Note that
\begin{align}
\Sigma^{WW}+(1-\lambda)\Sigma^{ZW}&=\dfrac{1}{2\beta\epsilon(1-\epsilon)}\left[2D^W\beta\epsilon+2\beta^2D^Z+(1-\lambda)(-2\beta(1-\epsilon)D^Z)\right]\notag\\
&=\dfrac{D^W}{1-\epsilon}\left[1+\beta-(1-\lambda)(1-\epsilon)\right]\notag\\
&>0,
\label{covariance factor}
\end{align}
where we used $D^Z:=\epsilon D^W$ in the second line and used the fact that $0\le\lambda\le1$ and $0<\epsilon<1$ in the last inequality.
From this relation and the fact that $\det\Sigma>0$, we conclude that
\begin{align}
\dot{I}^{Z\rightarrow W}_\lambda&>0\quad\text{and}\quad\dot{I}^{W\rightarrow Z}_\lambda<0
\end{align}
for all $\lambda\in[0,1]$.
\end{widetext}

\begin{figure}[t]
\center
\includegraphics[width=8.6cm]{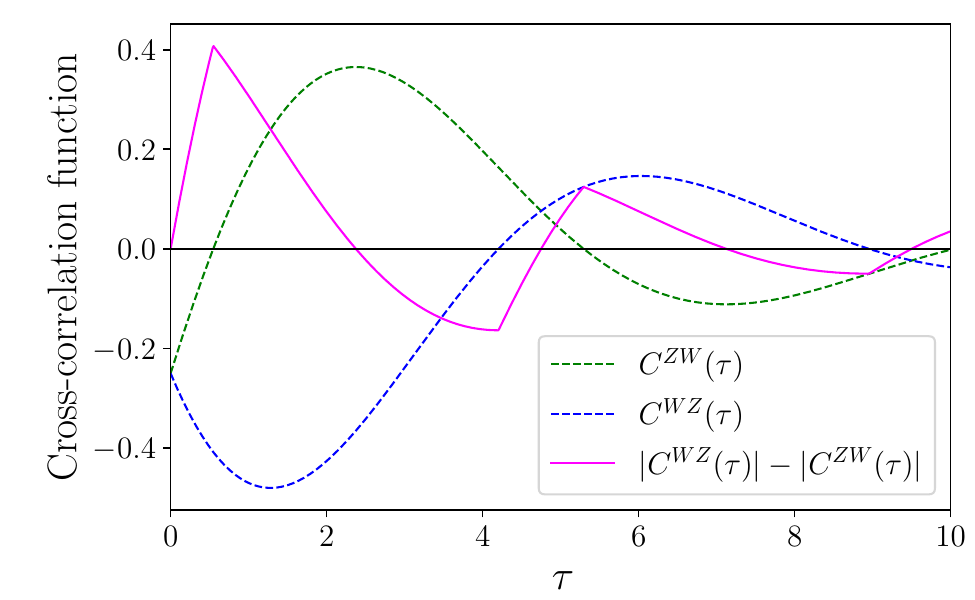}
\caption{$\tau$ dependence of $C^{WZ}(\tau)$, $C^{ZW}(\tau)$, and $|C^{WZ}(\tau)|-|C^{ZW}(\tau)|$. The parameter values are $\epsilon=D^W=1/2$, $\beta=1$. The characteristic period is given by $T=2\pi/\sqrt{\epsilon \beta-(1-\epsilon)^2/2}\simeq10$.}
\label{fig:asymmetric_cross_correlation}
\end{figure}
\begin{figure}[t]
\center
\includegraphics[width=8.6cm]{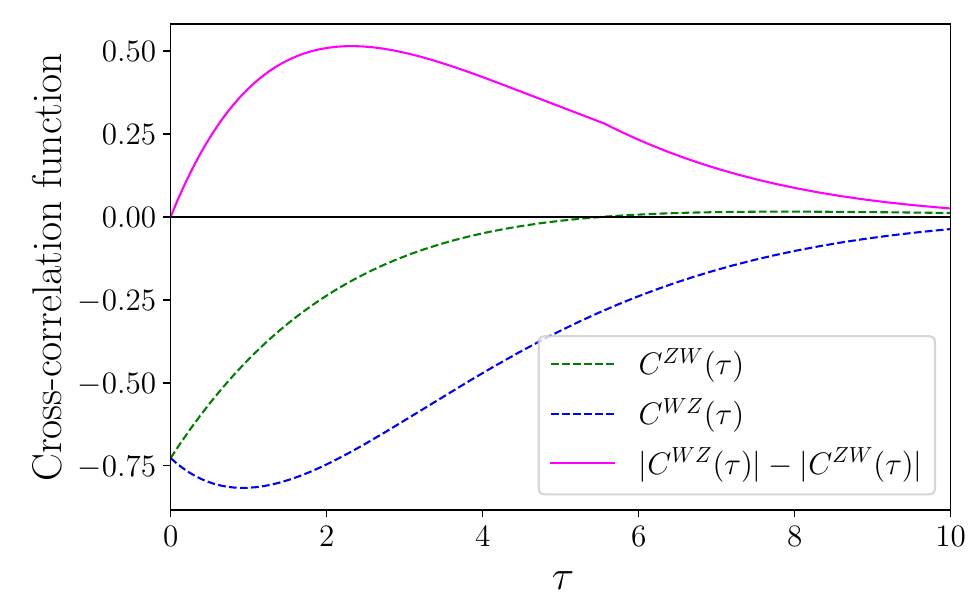}
\caption{$\tau$ dependence of $C^{WZ}(\tau)$, $C^{ZW}(\tau)$, and $|C^{WZ}(\tau)|-|C^{ZW}(\tau)|$. The parameter values are $\epsilon=0.2$, $D^W=1/2$, $\beta=1$. In this case, the resonance condition $\epsilon \beta-(1-\epsilon)^2/2>0$ is not satisfied.}
\label{fig:absolute_asymmetric_cross_correlation_epsilon_02}
\end{figure}

\section{Comparison with cross-correlation function\label{Comparison with cross-correlation function}}
Here, we provide the results of the cross-correlation function.
The cross-correlation function between turbulence $w$ and zonal flow $z$ at time lag $\tau$ is defined by
\begin{align}
C^{WZ}(\tau):=\dfrac{\langle (w_\tau-\langle w\rangle)(z_0-\langle z\rangle)\rangle}{\sqrt{\Sigma^{ZZ}\Sigma^{WW}}}.
\end{align}
This function is invariant under shifting of time in the steady state.
$C^{ZW}(\tau)$ is defined in a similar manner.
We also investigate the difference $|C^{WZ}(\tau)|-|C^{ZW}(\tau)|$, which quantifies the directional influence between turbulence and zonal flow.
The $\tau$-dependence of $C^{WZ}(\tau)$, $C^{ZW}(\tau)$, and the difference $|C^{WZ}(\tau)|-|C^{ZW}(\tau)|$ is shown in Figs.~\ref{fig:asymmetric_cross_correlation} and~\ref{fig:absolute_asymmetric_cross_correlation_epsilon_02}.

In Fig.~\ref{fig:asymmetric_cross_correlation}, the parameter values are the same as in Fig.~\ref{fig:four_pictures}(c) of the main text, where the resonance condition $\epsilon \beta-(1-\epsilon)^2/2>0$ is satisfied with the characteristic period $T=2\pi/\sqrt{\epsilon \beta-(1-\epsilon)^2/2}\simeq10$.
Note that the correlation function $C^{ZW}(\tau)$ exhibits a peak around $\tau\simeq T/4$, which is consistent with the quasi-cycles, where the fluctuations of zonal flow follow those of turbulence with a phase lag $\sim\pi/2$.
Notably, the difference $|C^{WZ}(\tau)|-|C^{ZW}(\tau)|$ is positive for a short time $\tau\lesssim T/4$, which suggests that turbulence is driven by zonal flow during this period.
Because the information flow also quantifies the directional influence during a short time period, this result is consistent with the positivity of the information flow $\dot{I}^{W\rightarrow Z}_\lambda$.

Figure~\ref{fig:absolute_asymmetric_cross_correlation_epsilon_02} shows the case where the resonance condition $\epsilon \beta-(1-\epsilon)^2/2>0$ is not satisfied.
Because there are no quasi-cycles in this case, the correlation functions do not oscillate.
Even in this case, the difference $|C^{WZ}(\tau)|-|C^{ZW}(\tau)|$ is positive for at least a short period of time, which is also consistent with the result of the information flow.

\bibliography{predator_prey}

\end{document}